\documentclass[%
 reprint,
 superscriptaddress,
 amsmath,amssymb,
 aps,
]{revtex4-2}

\usepackage{bbm}
\usepackage{dcolumn}
\usepackage{graphicx}

\usepackage{enumerate}
\usepackage{amsthm}
\usepackage{mathrsfs}
\usepackage{quantikz}

\begin{document}
\preprint{APS/123-QED}

\title{An Improved QFT-Based Quantum Comparator and Extended Modular Arithmetic Using One Ancilla Qubit}

\author{Yewei Yuan}
\affiliation{Origin Quantum Computing, Hefei, China}
\author{Chao Wang}
\affiliation{Origin Quantum Computing, Hefei, China}
\author{Bei Wang}
\affiliation{Origin Quantum Computing, Hefei, China}
\author{Zhao-Yun Chen}
\affiliation{Institute of Artificial Intelligence, Hefei Comprehensive National Science Center, Hefei, China}
\author{Meng-Han Dou}
\affiliation{Origin Quantum Computing, Hefei, China}
\author{Yu-Chun Wu}
\affiliation{CAS Key Laboratory of Quantum Information, University of Science and Technology of China, Hefei, 230026, China}
\affiliation{CAS Center for Excellence and Synergistic Innovation Center in Quantum Information and Quantum Physics,
University of Science and Technology of China, Hefei, 230026, China}
\affiliation{Hefei National Laboratory, University of Science and Technology of China, Hefei 230088, China}
\affiliation{Institute of Artificial Intelligence, Hefei Comprehensive National Science Center, Hefei, China}
\email{wuyuchun@ustc.edu.cn}
\author{Guo-Ping Guo}
\affiliation{CAS Key Laboratory of Quantum Information, University of Science and Technology of China, Hefei, 230026, China}
\affiliation{CAS Center for Excellence and Synergistic Innovation Center in Quantum Information and Quantum Physics,
University of Science and Technology of China, Hefei, 230026, China}
\affiliation{Hefei National Laboratory, University of Science and Technology of China, Hefei 230088, China}
\affiliation{Institute of Artificial Intelligence, Hefei Comprehensive National Science Center, Hefei, China}
\affiliation{Origin Quantum Computing, Hefei, China}
\email{gpguo@ustc.edu.cn}

\date{\today}
             

\begin{abstract}
Quantum comparators and modular arithmetic are fundamental in many quantum algorithms. Current research mainly focuses on operations between two quantum states. However, various applications, such as integer factorization, optimization, option pricing, and risk analysis, commonly require one of the inputs to be classical. It requires many ancillary qubits, especially when subsequent computations are involved. In this paper, we propose a quantum-classical comparator based on the quantum Fourier transform (QFT). Then we extend it to compare two quantum integers and modular arithmetic. Proposed operators only require one ancilla qubit, which is optimal for qubit resources. We analyze limitations in the current modular addition circuit and develop it to process arbitrary quantum states in the entire $n$-qubit space. The proposed algorithms reduce computing resources and make them valuable for Noisy Intermediate-Scale Quantum (NISQ) computers.

\end{abstract}

\maketitle


\section{\label{sec:level1}Introduction}
In recent years, quantum computers and related algorithms have been rapidly developing, driven by the laws of quantum mechanics and enhanced computing power. They can potentially solve many mathematical problems difficult for classical computers efficiently. The well-known Shor's algorithm\cite{shor1994algorithms} can solve hard problems such as integer factorization or discrete logarithms in polynomial time. Grover's algorithm\cite{grover1997quantum} can speed up an unstructured search problem quadratically. Quantum arithmetic is essential for these quantum algorithms. The main arithmetic includes comparison, basic, and modular arithmetic operations. It is worth noting that comparison is fundamental to modular arithmetic operations. 

Comparing a quantum state with a fixed classical integer can be seen in many algorithms and applications. While completing the comparison, most algorithms require preserving the original quantum input for subsequent calculations. For example, in finance, it can be used to compare the underlying price with the exercise price\cite{stamatopoulos2020option,ramos2021quantum,tang2021quantum,zhuang2023quantum} in derivatives pricing, calculate quantiles during VaR estimation\cite{woerner2019quantum,egger2020credit} and statistical arbitrage algorithm\cite{zhuang2022quantum}. Furthermore, the quantum-classical integer comparator can perform a bit or phase flip when the quantum state meets certain conditions. It is commonly seen in optimization and conditional search algorithms, especially in using Grover's algorithm. Recently, quantum image processing has focused more on algorithms that directly compare two quantum integers\cite{xia2018efficient,xia2019novel,li2020efficient,orts2021optimal}. The computational resources consumed by these quantum arithmetic circuits directly affect the efficiency of the entire algorithm. It is particularly important in the noisy intermediate scale quantum(NISQ) era.

\begin{figure*}[htbp]
    \centering
\begin{tikzpicture}
\node[scale=0.75] {
\begin{tikzcd}
\centering
\lstick{\ket{j_{1}}}&\gate{H}&\gate{R_{2}}&\qw&\cdots&&\gate{R_{n-1}}&\gate{R_{n}}&\qw&\qw&\qw&\qw&\qw&\qw&\qw&\qw&\qw&\qw&\qw&\qw&\qw&&&&&&&\lstick{\ket{0}+$e^{2\pi i0.j_1\cdots j_n}$\ket{1}}\\
\lstick{\ket{j_{2}}}&\qw&\ctrl{-1}&\qw&\qw&\qw&\qw&\qw&\gate{H}&\qw&\cdots&&\gate{R_{n-2}}&\gate{R_{n-1}}&\qw&\cdots&&\qw&\qw&\qw&\qw&&&&&&&\lstick{\ket{0}+$e^{2\pi i0.j_2\cdots j_n}$\ket{1}}\\
&\vdots&&\vdots\\
\lstick{\ket{j_{n-1}}}&\qw&\qw&\qw&\qw&\qw&\ctrl{-3}&\qw&\qw&\qw&\qw&\qw&\ctrl{-2}&\qw&\qw&\cdots&&\gate{H}&\gate{R_{2}}&\qw&\qw&&&&&&&\lstick{\ket{0}+$e^{2\pi i0.j_{n-1} j_n}$\ket{1}}\\
\lstick{\ket{j_{n}}}&\qw&\qw&\qw&\qw&\qw&\qw&\ctrl{-4}&\qw&\qw&\qw&\qw&\qw&\ctrl{-3}&\qw&\cdots&&\qw&\ctrl{-1}&\gate{H}&\qw&&&&&&\lstick{\ket{0}+$e^{2\pi i0.j_n}$\ket{1}}
\end{tikzcd}
};
\end{tikzpicture}
    \caption{The quantum circuit of QFT. Here $R_k$ represents the unitary operation of phase rotation by the angle $\frac{2\pi}{2^k}$. The number of CR operations required for the QFT on $n$ qubits input is $n(n+1)/2$, and the time complexity is $O(n^2)$.}
    \label{fig:lable001}
\end{figure*}
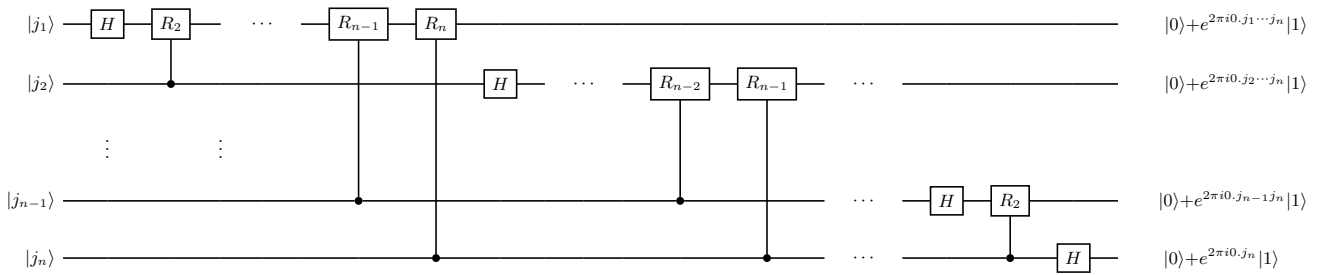

So far, two main common methods for constructing quantum arithmetic exist. They are based on Toffoli gates and quantum Fourier transform (QFT), respectively. For the first method, Vedral et al.\cite{vedral1996quantum} first give a carry-based version of modular adders, multipliers, and exponentiation circuits. Then, Gossett\cite{gossett1998quantum} borrowed the classical ``carry-save'' technique to design efficient arithmetic elements for Shor's algorithm. Cuccaro et al.\cite{cuccaro2004new} presented a linear-depth ripple-carry quantum adder circuit requiring only one ancilla qubit. H{\"a}ner et al.\cite{haner2016factoring} improved the Toffoli-based modular adder using $n$ dirty ancilla qubits. It can implement Shor's algorithm with $2n+2$ qubits. The method using QFT was first introduced by Draper\cite{draper2000addition} to reduce the number of carry qubits. Beauregard\cite{beauregard2002circuit} constructed the circuit for Shor's algorithm with $2n+3$ qubits using QFT. Other researchers\cite{ruiz2017quantum,csahin2020quantum,pavlidis2021quantum} proposed modifications to existing QFT-based arithmetic to improve their efficiency. 

In most quantum operations, QFT brings efficiency in many ways, especially in terms of flexible use of resources. When dealing with arithmetic operations between quantum states and given classical inputs, QFT-based arithmetic operations do not require additional qubits. Instead, classical information is loaded into phase rotation gates directly. Furthermore, the added phase can directly reflect a given linear transformation operation in the quantum state. Therefore, some carry-based MAJ-UMA operation circuits can be saved. Despite these advances, the design of quantum integer comparator remains challenging. Although the method comparing a classical integer through QFT-based subtracting is mentioned in \cite{beauregard2002circuit}, changing the original quantum state makes the quantum input unsuitable for subsequent calculations. So most approaches\cite{cuccaro2004new,takahashi,haner2016factoring,stamatopoulos2020option,ramos2021quantum} still require $n$ ancilla qubits with Toffoli gates. Therefore, by making the most of the QFT advantages, this paper attempts to minimize the qubits resource required for quantum comparison operations and the related modular arithmetic using QFT.

This paper proposes a novel quantum integer comparator based on QFT and analyses its performance. It can be extended to compare two
quantum integers and modular arithmetic. Unlike many previous works discussing arithmetic operations between two quantum states, our circuits are specifically designed for mixed input of quantum and classical information. Our proposed integer comparator reduces the ancillary qubits from $n+1$ to only one, which carries the comparison result. 
The depth of the circuit is 4 times of the QFT circuit.
The extended QFT modular addition circuit requires only one ancilla qubit without any ``dirty'' qubit and Toffoli gate. The modular addition circuit depth equals 8 QFTs. The quantum resource for modular addition is optimal. A circuit implementing Shor's algorithm can be further constructed using $2n+2$ qubits. The depth and size of the circuit are $O(n^3)$ and $O(n^3 \log{n})$. We also analyze the completeness applicability conditions of our proposed modular arithmetic operation and improve the algorithm for the complete operation on the entire $n$-qubit space.

The rest of this paper is organized as follows: Section II provides a brief overview of quantum arithmetic operations with quantum Fourier transform. Section III proposes the quantum integer comparator based on QFT and analyzes its performance. Then extends it to compare two quantum integers. Section IV extends the comparator to modular arithmetic and explores the circuit depth. After analysing the applicability conditions of our proposed adder, we improve it for the operation of the entire $n$-qubit space. Finally, Section V concludes this paper and discusses the potential applications of our proposed algorithms in various fields.

\section{\label{sec:level1}Preliminaries}
In this section, we briefly introduce the preliminaries 
of quantum Fourier transform (QFT) and describe the basic principles of QFT implementation of quantum arithmetic operations. There has been a significant amount of research\cite{cleve2000fast,nam2020approximate,park2022reducing} on circuit optimization for quantum Fourier transform and approximate QFT, which aims to reduce the computational resources like T-count and T-depth. However, this article does not focus on circuit optimization for QFT itself. Since the main component of our algorithm only involves QFT operations, we will calculate the complexity of our algorithm by treating QFT as a single unit while acknowledging that the actual circuit complexity depends on the method used to implement QFT.

\subsection{\label{sec:level2}Quantum Fourier Transform}
A vector $(x_{0},x_{1},\cdots,x_{N-1})$ can be transformed to a vector $(y_{0},y_{1},\cdots,y_{N-1})$ by the discrete Fourier inverse transform (IDFT) by the following equation:
\begin{equation}
    y_k\mapsto\dfrac{1}{\sqrt{N}}\sum\limits_{j=0}^{N-1}x_je^{2\pi ijk/N}.
\end{equation}

Similarly, the quantum Fourier transform (QFT), as a quantum implementation of the IDFT, transforms the amplitude of quantum states:
\begin{equation}
    |X\rangle=\sum_{j=0}^{N-1}x_{j}|j\rangle \longmapsto |Y\rangle=\sum_{k=0}^{N-1}y_{k}|k\rangle,
\end{equation}
where $y_{k}=\dfrac{1}{\sqrt{N}}\sum\limits_{j=0}^{N-1}e^{2\pi ijk/N}x_{j}$, $N = 2^n$, $n$ is the number of qubits, and the bases $|0\rangle, \cdots, |N-1\rangle$ are the computational bases of $n$ qubits.

If we consider the transformation on a single basis state, we can obtain the following form of QFT
\begin{equation}\label{test}
    |j\rangle\mapsto\dfrac{1}{\sqrt{N}}\sum\limits_{k=0}^{N-1}e^{2\pi ijk/N}|k\rangle, 
\end{equation}
and of inverse Fourier inverse transform(IQFT):
\begin{equation}
    |x\rangle\stackrel{\text{IQFT}}{ \longrightarrow} \dfrac{1}{\sqrt{N}}\sum\limits_{k=0}^{N-1} e^{-2\pi ixk/N}|k\rangle.
\end{equation}

The quantum circuit of the QFT application to the $n$-qubit state $|j\rangle$ is shown in Fig.~\ref{fig:lable001}. It is worth noting that when $|j\rangle = |0\rangle$, the transformation is equivalent to a series of Hadamard gates acting on each qubit individually.

\subsection{\label{sec:level2}QFT Arithmetic}
According to Eq.~(\ref{test}), the QFT transforms basis vector $|j\rangle$ into the power of amplitude $e^{2\pi ij/N}$. With the help of QFT and its inverse transformation, data can be transformed between the basis state and the amplitude. It is precisely this characteristic that makes QFT capable of flexibly handling arithmetic operations on the basis state. In arithmetic operations, it is customary to convert the computational basis to the so-called Fourier basis using QFT first, then perform the operation on the Fourier basis using phase and controlled phase gates, and then restore the state to the computational basis using IQFT finally. 

When using the Fourier basis to represent an integer $a$ on $n$ qubits, there will be a corresponding phase $\frac{2\pi a}{2^k}$ on the $k$-th qubit which is identical to the result of performing the quantum Fourier transform on $|a\rangle$. When we perform an arithmetic operation $a\mapsto f(a)$ in the Fourier basis, we are performing an operation from the phase sequence $\{\frac{2\pi a}{2^k}\}$ to the phase sequence $\{\frac{2\pi f(a)}{2^k}\}$. 

\begin{figure}[h]
    \centering
\begin{tikzpicture}
\node[scale=0.9] {
\begin{tikzcd}
\centering
\lstick{\ket{j}}&\qwbundle
{n}&\gate[wires=1, style={yshift=0pt}][1.3cm][1.3cm]{QFT_{n}}&\gate[wires=1,style={yshift=0pt
}][1.3cm][1.3cm]{\begin{array}{cc} \bigotimes U_{k} (\frac{2\pi a}{2^{n-k}}) \end{array}}&\gate[wires=1, style={yshift=0pt}][1.3cm][1.3cm]{QFT^{\dagger}_{n}}&\qw&&&\lstick{\ket{j+a}}\\
\end{tikzcd}
};
\end{tikzpicture}
    \caption{A simple circuit adding a number to a register using QFT for principle description. Due to the periodicity of the phase, results outside the interval $[0, N)$ will overflow and fall back within $[0, N)$.}
    \label{fig:lable002}
\end{figure}
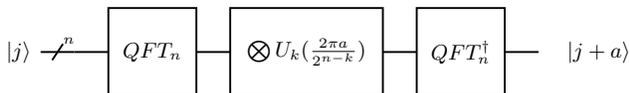

Taking the example of adding a number to a register, we need to implement $|j\rangle \mapsto |j+a\rangle$. The procedure to implement this operation is the following:
\begin{equation}
\begin{aligned}
	|j\rangle_{n} 
    &\stackrel{\text{QFT$_n$}}{\xrightarrow{\qquad}} \dfrac{1}{\sqrt{N}}\sum\limits_{k=0}^{N-1}e^{2\pi ijk/N}|k\rangle,\\
    &\stackrel{\bigotimes\text{U}_k}{\xrightarrow{\qquad}}\dfrac{1}{\sqrt{N}}\sum\limits_{k=0}^{N-1}e^{2\pi i(j+a)k/N}|k\rangle,\\
    &\stackrel{\text{IQFT}_{n}}{\xrightarrow{\qquad}} |j+a\rangle_{n} ,
 \end{aligned}
 \end{equation}
where $\bigotimes U_k$ represents rotating the $k$-th qubit by the angle in the phase sequence $\{\frac{2\pi a}{2^k}\}$ using the phase rotation gates. It should be noted that after QFT, the qubits are reversed in order, so in practical operations, if no swap gates are used, the original register actually performs the phase rotation $\{\frac{2\pi a}{2^{n-k}}\}$. The quantum circuit is shown in Fig.~\ref{fig:lable002}.

\section{\label{sec:level1}Quantum Comparator}
\subsection{\label{sec:level2}Quantum Integer Comparator} \label{qic}
Unlike most of the literature on quantum arithmetic, our focus here is not on directly comparing two quantum states but on the size relationship between a quantum state and a given classical number. The operation that the quantum integer comparator needs to implement is to compare the computational basis state $|j\rangle_n$ with a given classical integer $a$ of a fixed value and flip the target qubit when $j$ is less than (or greater than) $a$. Here we use the case of less than in the following. The quantum circuit needs to ensure that the basis state of the original register $|j\rangle_n$ remains unchanged before and after the operation: 
\begin{equation}
    |j\rangle_n|0\rangle \rightarrow |j\rangle_n|j < a\rangle.
\end{equation}
We need to note that here $a$ is assumed to be within the range $[0, N)$, where $N=2^n$, since in other cases the comparison result is certain. 

The most commonly used approach for quantum integer comparator is to modify the details of the ripple-carry adder based on the circuit proposed by Cuccaro et al.\cite{cuccaro2004new} and fully utilize the classical bit information on the circuit. Further modifications can be found in \cite{stamatopoulos2020option,ramos2021quantum}. In addition to the qubit representing the comparison result, $n$ auxiliary qubits are needed for recording the carry results of all bits. It requires $n-1$ Toffoli gates and 2 CNOT gates in the circuit. This method can perform reverse calculations on the ancilla qubits. If ancilla qubits need a reset, the complete circuit requires $2n-2$ Toffoli and 3 CNOT gates on the total $2n+1$ qubits.

\begin{figure*}[htb]
\begin{tikzpicture}
\node[scale=0.85] {
\begin{tikzcd}[thin lines]
\centering
\lstick{\ket{x}}&\qwbundle
{n}&\gate[wires=2, disable auto height]{QFT_{n+1}}\gategroup[2,steps=3,style={
rounded corners,fill=blue!20, inner xsep=0pt, inner ysep=0pt},
background]{} &\gate[wires=2, disable auto height]{\begin{array}{cc} \bigotimes U_{k} (-\frac{2\pi a}{2^{n+1-k}}) \end{array}}&\gate[wires=2, disable auto height]{QFT^{\dagger}_{n+1}}&\gate[wires=1, style={yshift=-10pt}][1.3cm][1.3cm]{QFT_{n}}\gategroup[1,steps=3,style={
rounded corners,fill=green!20, inner xsep=0pt, inner ysep=10pt, yshift=-10pt},
background]{}&\gate[wires=1,style={yshift=-10pt
}][1.3cm][1.3cm]{\begin{array}{cc} \bigotimes U_{k}(\frac{2\pi a}{2^{n-k}}) \end{array}}&\gate[wires=1, style={yshift=-10pt}][1.3cm][1.3cm]{QFT^{\dagger}_{n}}&\qw&&\lstick{\ket{x}}\\[1cm]
\lstick{\ket{0}}&\qw&\qw&\qw&\qw&\qw&\qw&\qw&\qw&&&\lstick{\ket{x<a}}
\end{tikzcd}
};
\end{tikzpicture}
\caption{The proposed QFT-based quantum integer comparator. The comparison part is over blue, and the reset part is over green.}
\label{fig:lable003}
\end{figure*}
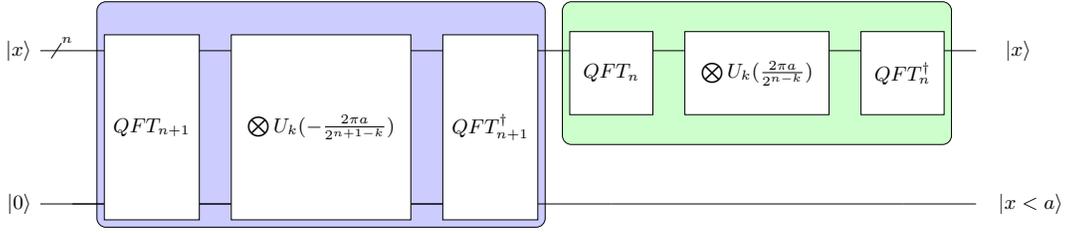

The proposed comparator in Fig.~\ref{fig:lable003} can achieve this operation by using one QFT adder as a comparison part and the other as a reset part. No additional auxiliary qubits except the qubit carrying the comparison result are needed.

The first part follows the comparison operation proposed in \cite{beauregard2002circuit}. For an input quantum state $|x\rangle_n$, it uses QFT to perform a subtraction between $x$ and $a$ on the Fourier basis and obtains the comparison information. Here, $|x\rangle_n$ can be regarded as a computational basis or superposition state. 
\begin{equation}\label{test3}
    |x\rangle_n|0\rangle \rightarrow |x-a\rangle_{x\geq a}|0\rangle + |x-a+N\rangle_{x<a}|1\rangle.
\end{equation}

After the first adder, the state $|x\rangle|0\rangle$ comes to $|x-a\rangle_{x\geq a}|0\rangle + |x-a+N\rangle_{x<a}|1\rangle$. Then we use the second adder to reset $|x\rangle$ through a $n$-local addition to $a$:

\begin{equation}\label{test2}
    \begin{aligned}[b]
        &|x-a\rangle_{x\geq a}|0\rangle + |x-a+N\rangle_{x<a}|1\rangle \\
        &\hspace{0em}
        \begin{aligned}[t]
        &\stackrel{+a}{\xrightarrow{\qquad}}|x-a+a\rangle_{x\geq a}|0\rangle \\
        & + |(x-a+N+a)\bmod N\rangle_{x<a}|1\rangle \\
        \end{aligned}\\
        &= |x\rangle|x<a\rangle.
    \end{aligned}
\end{equation}

It is worth noting that no comparison information is required for resetting, as the local addition on the Fourier basis itself includes the modulo operation concerning $N$. 
Finally, we obtained the state $|x\rangle|x<a\rangle$ according to Eq.(\ref{test2}) and completed the entire comparison work. For other relations like less than or equal to and greater than or equal to, etc., we can take $a = a + 1$ and apply a NOT gate, respectively.
Our proposed integer comparator uses 4 QFT(IQFT) operations containing
$2n^2$ CR gates for $n$ inputs and no other multi-qubit gates. If evaluated in terms of CNOT gates, the number should be multiplied by 2. 

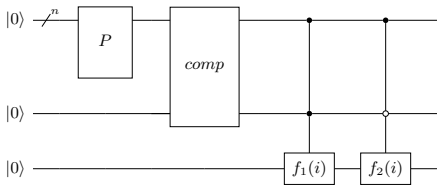
\begin{figure}[h]
    \centering
\begin{tikzpicture}
\node[scale=0.7] {
\begin{tikzcd}[thin lines]
\centering
\lstick{\ket{0}}&\qwbundle{n}&\gate[wires=1,style={yshift=-12pt, inner xsep=4pt,  inner ysep=10pt}]{\begin{array}{cc} P \end{array}}&\qw&\gate[wires=2,style={inner xsep=4pt}, disable auto height]{comp}&\qw&\ctrl{1}&\ctrl{1}&\qw\\[0.7cm]
\lstick{\ket{0}}&\qw&\qw&\qw&\qw&\qw&\ctrl{1}&\octrl{1}&\qw\\
\lstick{\ket{0}}&\qw&\qw&\qw&\qw&\qw&\gate{f_{1}(i)}&\gate{f_{2}(i)}&\qw
\end{tikzcd}
};
\end{tikzpicture}
    \caption{The construction of piecewise function circuits. Block P, comp, $f_1$ and $f_2$ represent the probability distribution loading operation, comparator, and the function Oracle for different conditions, respectively. This circuit is frequently seen in applications using quantum Monte Carlo integration such as option pricing.}
    \label{fig:lable099}
\end{figure}

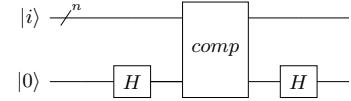
\begin{figure}[h]
    \centering
\begin{tikzpicture}
\node[scale=0.85] {
\begin{tikzcd}[thin lines]
\centering
\lstick{\ket{i}}&\qwbundle{n}&\qw&\gate[wires=2, disable auto height]{comp}&\qw&\qw\\
\lstick{\ket{0}}&\qw&\gate{H}&\qw&\gate{H}&\qw
\end{tikzcd}
};
\end{tikzpicture}
    \caption{A less-than/greater-than oracle flips the phase based on the comparison result. The original quantum state $|i\rangle$ should remain unchanged.}
    \label{fig:lable100}
\end{figure}

This operation appears in various algorithms involving threshold comparison and thus has a wide range of applications, such as optimization, search, and computation with piecewise functions. When completing quantum Monte Carlo integration and other algorithms, if the target function is piecewise, we need many comparison operations to perform different operations on the sequentially controlled quantum states in different intervals, which requires repeatedly calling the comparator and ensuring that the original quantum register remains unchanged. This process can be seen in Fig.~\ref{fig:lable099}. 
In the amplitude amplification operation in the Grover algorithm, if we perform a phase flip on the quantum states marked in a specified range as shown in Fig.~\ref{fig:lable100}, we also need to perform comparison operations while ensuring that the original quantum state remains unchanged. In addition, constructing constraints in optimization problems is also a functionality that quantum integer comparator can achieve.

\subsection{\label{sec:level2}Quantum States Comparator}
For comparators between two quantum states, an original general-purpose comparator based on subtraction was proposed in \cite{oliveira2007quantum}. Wang et al.\cite{wang2012design} proposed a comparison scheme that requires $2n$ auxiliary qubits. Engin Sahin\cite{csahin2020quantum} designs a quantum comparator requiring $(3n^2 + 9n + 82)/2$ operations for $n \times n$ inputs. However, this type of comparator based on QFT subtraction can cause a change in the quantum state of an original register. The method proposed later in this paper can effectively address this issue. The research \cite{xia2018efficient,xia2019novel,li2020efficient,orts2021optimal} continuously optimized the computational resources of quantum comparators and can use only one ancillary qubit, but these are all based on carry operations. 

In our work, the quantum states comparator can be easily extended from the proposed integer comparator. The circuit is constructed similarly to Draper adder\cite{draper2000addition}, with the integer $a$ replaced by register $|x_1\rangle_n$. The other register $|x_2\rangle_n$ can be compared by performing controlled phase rotations using each qubit of $|x_1\rangle_n$ in the decomposition as control. The rotation angles follow the binary expansion of $|x_1\rangle_n$. We leave the detail in Appendix \ref{appendix}. 

Compared with the current QFT quantum comparator based on the QFT subtractor, we can ensure that the quantum states being compared remain unchanged before and after the circuit and obtain the comparison result without obtaining the subtraction result. Compared with the commonly used carry-based comparators, the proposed quantum comparator require no extra ancillary qubit. Furthermore, supposing a given linear transformation, such as comparing $ax_1+b$ and $x_2$, is present in the quantum state being compared. In that case, it can be directly reflected in the additional phase on the Fourier basis. There is no need to modify the compared quantum state itself with extra  quantum arithmetic operations. 

Compared to the quantum integer comparator, the quantum state comparator requires extra $2n^2$ CR gates for controlled phase addition, in addition to the original 4 QFT(IQFT) circuits. The number of operations required for the
proposed states comparator is $4n^2$ for $n\times n$ inputs if we count the number of CR gates.

\section{\label{sec:level1}Modular Arithmetic}
Modular arithmetic is always associated with comparison operations. It is natural to consider using the comparator in \ref{qic} to derive a modulo operation for a given fixed integer and then optimise the existing modular arithmetic circuit.

Firstly, discussing the prerequisite restrictions that make the subsequent operation meaningful is essential. Specifically, the given integer $M$ that we want to take the modulus of must belong to the range $[N/2, N)$. Since values of $M$ greater than $N$ would result in some outcomes that the given quantum register cannot represent and instead must be expressed as an overflow value of $N$. On the other hand, values of $M$ smaller than $N/2$ would also lead to many overflow issues of $M$. Therefore, we need to assume that the number of qubits, $n$, is the minimum number representing integer $M$, which is typically the case for most applications. Moreover, it is worth noting that, unlike the modular arithmetic required in Shor's algorithm, when constructing the circuit, we consider the input quantum state $|x\rangle$ to be a more general superposition state rather than a specific computational basis state.

\subsection{\label{sec:level2}Quantum Modular Operation}\label{qmo}
According to Eq.(\ref{test3}), the state after the first stage of the integer comparator in Fig.~\ref{fig:lable003} is:
\begin{equation}
    |x-a\rangle_{x\geq a}|0\rangle + |x-a+N\rangle_{x<a}|1\rangle.
\end{equation}
Using the comparison result, we can reset the state when $x$ is less than $a$ and obtain the state for $x$ modulo $a$. Specifically, we need to obtain the state
\begin{equation} \label{test4}
    |x-a\rangle_{x\geq a}|0\rangle + |x\rangle_{x<a}|1\rangle.
\end{equation}
To accomplish this, we perform phase rotations using CR gates to complete the Fourier-based addition $(x-a+N) + (a-N) = x$ if the comparator result qubit is in the state of $|1\rangle$. 

\begin{figure}[h]
    \centering
\begin{tikzpicture}
\node[scale=0.59] {
\begin{tikzcd}[thin lines]
\centering
\lstick{\ket{x}}&\qwbundle
{n}&\gate[wires=2, disable auto height]{QFT_{n+1}}&\gate[wires=2, disable auto height]{\begin{array}{cc} \varphi_{n+1} \\[0cm] (-M)  \end{array}}&\gate[wires=2, disable auto height]{QFT^{\dagger}_{n+1}}&\gate[wires=1, style={yshift=-11pt}][1.3cm][1.3cm]{QFT_{n}}&\gate[wires=1, style={yshift=-11pt}][1.3cm][1.3cm]{\varphi_{n}(M)}&\gate[wires=1, style={yshift=-11pt}][1.3cm][1.3cm]{QFT^{\dagger}_{n}}&\qw&&&&\lstick{\ket{x\bmod\ M}}\\[1cm]
\lstick{\ket{0}}&\qw&\qw&\qw&\qw&\qw&\ctrl{-1}&\qw&\qw&&&\lstick{\ket{x<M}}
\end{tikzcd}
};

\end{tikzpicture}
    \caption{The quantum circuit computing $x \bmod {M}$. Here we use $\varphi_n(M)$ represent phase addition on Fourier basis replacing previous representation $\bigotimes U_K$.}
    \label{fig:lable004}
\end{figure}
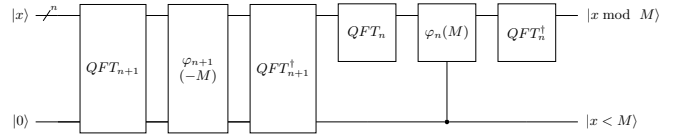

Since it is a Fourier basis on $n$ qubits and $N$ is the period, adding $a$ is equivalent to adding $a-N$. Here we take $a=M$. And then, we obtain the quantum circuit shown in Fig.~\ref{fig:lable004} for computing $x$ modulo $M$.

The circuit implement
\begin{equation}
    |x\rangle_n|0\rangle \rightarrow |x\bmod {M}\rangle|x<M\rangle.
\end{equation}
The qubit carrying the comparison information is not restored to $|0\rangle$ in the end. This is inevitable because the operators in quantum computing are unitary. Meanwhile, this circuit can be used to prepare an initial state with constraints. 

\subsection{\label{sec:level2}Modular Addition Operation}
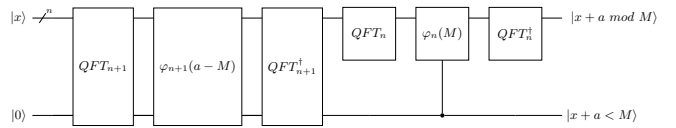
\begin{figure}[ht]
    \centering
\begin{tikzpicture}
\node[scale=0.54] {
\begin{tikzcd}[thin lines]
\centering
\lstick{\ket{x}}&\qwbundle
{n}&\gate[wires=2, disable auto height]{QFT_{n+1}}&\gate[wires=2, disable auto height]{\varphi_{n+1}(a-M)}&\gate[wires=2, disable auto height]{QFT^{\dagger}_{n+1}}&\gate[wires=1, style={yshift=-11pt}][1.3cm][1.3cm]{QFT_{n}}&\gate[wires=1,style={yshift=-11pt}][1.3cm][1.3cm]{\varphi_{n}(M)}&\gate[wires=1, style={yshift=-11pt}][1.3cm][1.3cm]{QFT^{\dagger}_{n}}&\qw&&&&&\lstick{\ket{x+a\ mod\ M}}\\[1cm]
\lstick{\ket{0}}&\qw&\qw&\qw&\qw&\qw&\ctrl{-1}&\qw&\qw&&&&\lstick{\ket{x+a<M}}
\end{tikzcd}
};
\end{tikzpicture}
    \caption{The quantum circuit computing $x+a\bmod {M}$ obtained directly by extending the previous modular circuit.}
    \label{fig:lable005}
\end{figure}

Thanks to the flexibility of calculations on the Fourier basis, if we further want to perform modular addition, we can add the given integer $a$ where $a$ is less than $M$ at the Fourier basis in the first part. Thus obtain the circuit in Fig.~\ref{fig:lable005}.

We can further analyze its applicability by examining the final result of the circuit under different comparison results in the first part of the circuit:

\begin{enumerate}
     \item $x+a-M\geq 0 \rightarrow x+a\geq M$ \label{item}\\
        Get $|x+a \bmod{M}\rangle$ if and only if $x+a-M<M$, i.e. \textbf{$x+a<2M$}\\
        This condition is automatically met if $x\in[0, M)$. But when we want to generalize it to all $|x\rangle$ in $n$-qubit space, we need $a$ to satisfy $a\leq 2M-N$ constantly. 
     \item $x+a-M<0 \rightarrow x+a<M$ \\
        Get $|x+a-M+N\rangle$ after the first stage since $M<N$. \\
        Finally, get $|x+a\rangle$, equal to $|x+a \bmod{M}\rangle$ since $x+a-M<0$.
\end{enumerate}

In summary, the above circuit can be seen as a modular addition circuit under the precondition constraint $x\in[0, M)$, or a modular addition circuit for all $x$ in the entire $n$-qubit space provided that the given fixed number $a$ satisfies the condition $a\leq2M-N$. The final state of the circuit is as follows:
\begin{equation}
    |x+a-M\rangle_{x+a\geq M}|0\rangle + |x+a\rangle_{x+a<M}|1\rangle
\end{equation}

Inspired by the original Beauregard adder\cite{beauregard2002circuit}, we can utilize the modular addition result in Eq.(\ref{test4}) to reset the ancillary qubit, allowing it to be reused under the constraint of a pre-existing constraint $x\in[0, M)$. This constraint also meets the previous applicability condition in \ref{item}. After the subtraction of $a$ on the Fourier basis, the quantum state becomes
\begin{equation}
    |x-M+N\rangle_{x+a\geq M}|1\rangle + |x\rangle_{x+a<M}|1\rangle
\end{equation}

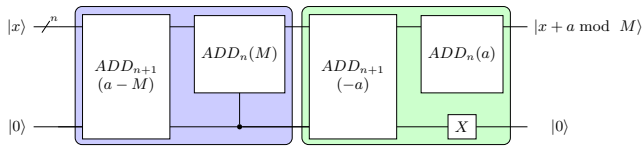
\begin{figure}[h]
    \centering
\begin{tikzpicture}
\node[scale=0.65] {
\begin{tikzcd}[thin lines]
\centering
\lstick{\ket{x}}&\qwbundle
{n}&\gate[wires=2, disable auto height]{\begin{array}{cc} ADD_{n+1}  \\[0cm] (a-M) \end{array}}\gategroup[2,steps=2,style={
rounded corners,fill=blue!20, inner xsep=1pt, inner ysep=0pt},
background]{}&\gate[wires=1,style={yshift=-16pt, inner xsep=0pt,  inner ysep=14pt}]{ADD_n(M)}&\gate[wires=2, disable auto height]{\begin{array}{cc} ADD_{n+1}  \\[0cm] (-a) \end{array}}\gategroup[2,steps=2,style={
rounded corners,fill=green!20, inner xsep=1pt, inner ysep=0pt},
background]{}&\gate[wires=1,style={yshift=-16pt, inner xsep=0pt,  inner ysep=14pt}]{ADD_n(a)}&\qw&&&&&\lstick{\ket{x+a\bmod\ M}}\\[1cm]
\lstick{\ket{0}}&\qw&\qw&\ctrl{-1}&\qw&\gate{X}&\qw&&\lstick{\ket{0}}
\end{tikzcd}
};
\end{tikzpicture}
    \caption{The proposed QFT-based quantum modular adder with one reusable ancilla qubit. The modular addition part is over blue background and the reset part is over green background. Here we adopt a more concise representation to denote addition finished by the Fourier basis. The circuit in the first part is identical to the one in Fig.~\ref{fig:lable005}.}
    \label{fig:lable006}
    
\end{figure}

At this point, adding the value $a$ to the first register can restore it to $x+a \bmod {M}$, and applying a CNOT gate to the second register can restore it to $|0\rangle$. The resulting circuit is shown in Fig.~\ref{fig:lable006}. When comput

                                           ing modular subtraction, it is equivalent to computing modular addition with $M-a$, which is still within the applicable range of the modular adder. 

This is currently the optimal quantum modular addition(subtraction) operation we have obtained for quantum input $|x\rangle$ and classical input $a$ where $a$ and $|x\rangle$ are promised to be integers mod $M$. The proposed adder only uses one resettable ancillary qubit. We used a total of 8 QFT (IQFT) operations and additional $n^2$ CR gates for the controlled phase addition. The number of operations required for the proposed adder is $5n^2$ for $n$ inputs if we count the number of CR gates. Compared with other modular adders, Beauregard\cite{beauregard2002circuit} requires 2 ancillary qubits, which leads to 1 more qubit to implement Shor's algorithm. Both Takahashi\cite{takahashi} and H{\"a}ner\cite{haner2016factoring} use the comparator-adder-comparator construction containing Toffoli-based oprations with $n$ reusable auxiliary qubits.

Furthermore, the construction methods for modular addition between quantum states, modular multiplication, and modular exponentiation can all be extended from the circuit of modular addition between quantum state and given integer. The extension method is relatively fixed and has been described in previous papers on implementing Shor's algorithm, including the work of Beauregard\cite{beauregard2002circuit}. We discuss the method in Appendix \ref{appendix2}. The depth and size of the implementation are $O(n^3)$ and $O(n^3 \log{n})$. Therefore, the proposed modular addition circuit can be used to complete a set of qubit-saving modular arithmetic operations for applications such as Shor's algorithm.

\begin{figure}[ht]
    \centering
\begin{tikzpicture}
\node[scale=0.5] {
\begin{tikzcd}[thin lines]
\centering
\lstick{\ket{0}}&\qw&\qw&\qw&\qw&\gate[wires=2, disable auto height]{\begin{array}{cc} ADD_{n+1}  \\[0cm] (a-M) \end{array}}&\ctrl{1}&\gate[wires=2, disable auto height]{\begin{array}{cc} ADD_{n+1}  \\[0cm] (-a) \end{array}}&\gate{X}&\qw&&\lstick{\ket{0}}\\[1cm]
\lstick{\ket{x}}&\qwbundle
{n}&\qw&\gate[wires=2, disable auto height]{\begin{array}{cc} ADD_{n+1}  \\[0cm] (-M) \end{array}}&\gate{\begin{array}{cc} ADD_{n}  \\[0cm] (M) \end{array}}&\qw&\gate{\begin{array}{cc} ADD_{n}  \\[0cm] (M) \end{array}}&\qw&\gate{\begin{array}{cc} ADD_{n}  \\[0cm] (a) \end{array}}&\qw&&&&&\lstick{\ket{x+a\ mod\ M}}\\[1cm]
\lstick{\ket{0}}&\qw&\qw&\qw&\ctrl{-1}&\qw&\qw&\qw&\qw&\qw&&&\lstick{\ket{x<M}}
\end{tikzcd}
};
\end{tikzpicture}
    \caption{The proposed QFT-based quantum modular adder for all $|x\rangle_n$ in $n$-qubit space.}
    \label{fig:lable007}
\end{figure}
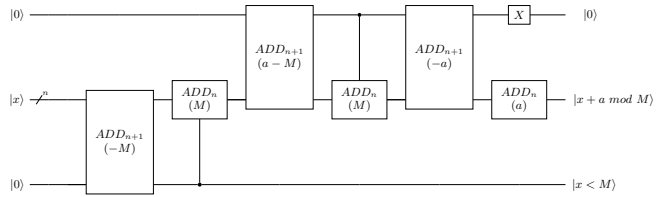

Finally, if we want to obtain a complete modular adder on n-qubits space that does not have any preconditions on $x$ or restrictions on the range $[0, 2M-N)$ of $a$, we can use the modular reduction circuit introduced earlier in \ref{qmo} to perform modular reduction on the state $|x\rangle$ before the modular addition operation. The quantum circuit is shown in Fig.~\ref{fig:lable007}. However, it should be noted that the auxiliary qubit introduced here is not reusable. Can we find a quantum circuit that meets the above requirements and uses auxiliary qubits that can all be reused? The answer is no, for the same reason as our discussion of the modulo operation in the \ref{qmo}. When $x>M$, it means that the same modulo value may correspond to two overflow indicators (overflow or not overflow) at the same time. If we want to reset the qubit that records the overflow information based on the modulo value at this time, this operation cannot be unitary. Therefore, it cannot be implemented through a quantum program.

\section{\label{sec:level1}Conclusions and Discussion}
In this work, we present a new quantum integer comparator based on the quantum Fourier transform and evaluate its performance. It can be widely used in algorithms with threshold comparison, optimization, conditional search, etc. The proposed algorithm is scalable and can be extended to modular arithmetic. Compared to previous works that directly focus on purely quantum arithmetic operations, our approach considers arithmetic operations that involve classical values and further generalizes to arithmetic operations between quantum states. Our proposed algorithms reduce the required computing resources compared to current quantum arithmetic algorithms and have potential applications in various fields. This is also highly beneficial for validating the algorithms through numerical experiments. 

The quantum resource for extended modular addition is optimal. The construction differs from current methods, including \cite{beauregard2002circuit, takahashi, haner2016factoring}. A circuit implementing Shor's algorithm can be further constructed using $2n+2$ qubits. We also analyze the completeness applicability conditions of our proposed modular arithmetic operation and improve the algorithm for the complete operation on the entire $n$-qubit space. Unlike the modular arithmetic operation required in Shor's algorithm, we have considered the input quantum state $|x\rangle$ as a more general superposition state rather than a specific computational basis state. We believe this is beneficial and can be applied to solve practical problems, such as using our circuit in combination with Grover's algorithm to find the inverse element of an element, which requires obtaining a uniform superposition state on the entire $n$-qubit space at the beginning.

We hope this work will provide new solutions to quantum arithmetic problems and further develop and improve the construction efficiency of Shor's algorithm and various Oracles in quantum algorithms.

\bibliography{reference}

\clearpage
\onecolumngrid
\appendix
\section{Quantum States Comparator Circuit}\label{appendix}
Like the greater-than and less-than signs in classical computing, comparing the sizes between quantum states is also indispensable. Many quantum algorithms rely on quantum comparators, with the simplest being to compare the sizes and filter out the states of the required qubits. Here we show details of comparator between $|x_{1}\rangle_{n}$ and $|x_{2}\rangle_{n}$, its similar with quantum state and integer value, use $|x_{1}\rangle_{n}$ control $U_{1}$ rotation rather than fixed angle. and the state changes below. Here, we need to control the quantum state of $|x_{1}\rangle_{n}$ to control the quantum state of $|x_{2}\rangle_{n}$, which requires traversing control, and this will take approximately $n^{2}$ gates. 

\begin{figure*}[ht]
    \centering
\begin{tikzpicture}
\node[scale=1] {
\begin{tikzcd}[thin lines]
\centering
\lstick{\ket{x_{1}}}&\qwbundle{n}&\qw&\ctrl{1}&\qw&\qw&\qw&\ctrl{1}&\qw&\qw&\qw&&\lstick{\ket{x_{1}}}\\
\lstick{\ket{x_{2}}}&\qwbundle{n}&\gate[wires=2, disable auto height]{QFT_{n+1}}&\gate[wires=2, disable auto height]{\begin{array}{cc}\bigotimes U_{k} (-\frac{2\pi x_1}{2^{n+1-k}})  \end{array}}&\gate[wires=2, disable auto height]{QFT^{\dagger}_{n+1}}&\gate[wires=1, style={yshift=-11pt}][1.3cm][1.3cm]{QFT_{n}}&\qw&\gate[wires=1,disable auto height, style={yshift=-10pt, inner xsep=22pt,  inner ysep=10pt}]{\begin{array}{cc} \bigotimes U_{k}(\frac{2\pi x_1}{2^{n-k}}) \end{array}}&\qw&\gate[wires=1, style={yshift=-11pt}][1.3cm][1.3cm]{QFT^{\dagger}_{n}}&\qw&&\lstick{\ket{x_{2}}}\\[1cm]
\lstick{\ket{0}}&\qw&\qw&\qw&\qw&\qw&\qw&\qw&\qw&\qw&&&\lstick{\ket{x_{2}<x_{1}}}
\end{tikzcd}
};
\end{tikzpicture}
    \caption{The proposed QFT-Based quantum comparator extended by the integer comparator.}
    \label{fig:lable010}
\end{figure*}
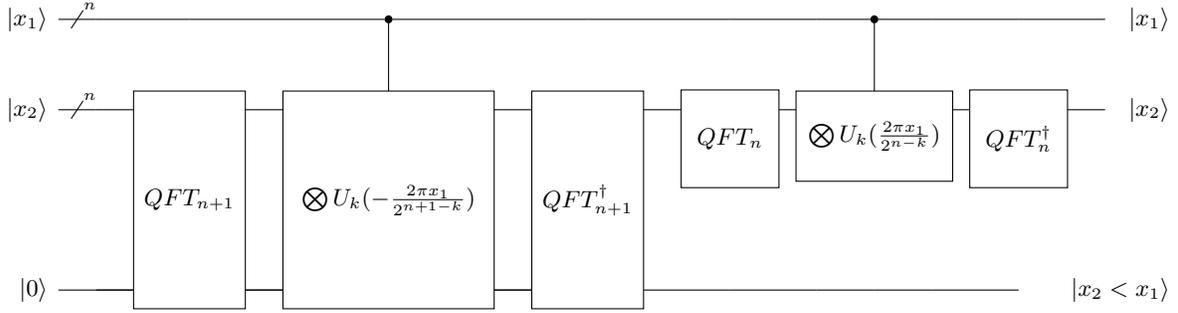

\begin{align}
    &|x_{1}\rangle|0\rangle|x_{2}\rangle \stackrel{I^{\otimes n}\otimes QFT_{n+1}}{\xrightarrow{\qquad}} \\
    &|x_{1}\rangle \sum\limits_{k=0}^{N-1}{e^{2\pi i x_{2}k/2N}|k\rangle}\stackrel{I^{\otimes n}\otimes U_{k}(-0.0x_{1})}{\xrightarrow{\qquad}}\\
    &|x_{1}\rangle\sum\limits_{k=0}^{N-1}{e^{2\pi i (x_{2}-x_{1})k/2N}|k\rangle}  \stackrel{equal}{\xrightarrow{\qquad}} \\
    &|x_{1}\rangle_{x_{2}\geq x_{1}}\sum\limits_{k=0}^{N-1}{e^{2\pi i (x_{2}-x_{1})k/2N}|k\rangle} + |x_{1}\rangle_{ x_{2}< x_{1}}\sum\limits_{k=0}^{N-1}{e^{2\pi i (2N +x_{2}-x_{1})k/2N}|k\rangle}
     \stackrel{I^{\otimes n}\otimes IQFT_{n+1}}{\xrightarrow{\qquad}}\\ 
    &|x_{1}\rangle|0\rangle_{x_{2} \geq x_{1}}|x_{2}-x_{1}\rangle +
    |x_{1}\rangle|1\rangle_{x_{2} < x_{1}}|N + x_{2}-x_{1}\rangle \stackrel{I^{\otimes n+1}\otimes QFT^{}_{n}}{\xrightarrow{\qquad}}\\
    &|x_{1}\rangle|0\rangle_{x_{2} \geq x_{1}}\sum\limits_{k=0}^{N-1}{e^{2\pi i (x_{2}-x_{1})k/N}|k\rangle} +
    |x_{1}\rangle|1\rangle_{x_{2} < x_{1}}\sum\limits_{k=0}^{N-1}{e^{2\pi i (N + x_{2}-x_{1})k/N}|k\rangle}\stackrel{I^{\otimes n+1}\otimes U_{k}(0.x_{1})}{\xrightarrow{\qquad}}\\
    &|x_{1}\rangle|0\rangle_{x_{2} \geq x_{1}}\sum\limits_{k=0}^{N-1}{e^{2\pi i x_{2}k/N}|k\rangle} +
    |x_{1}\rangle|1\rangle_{x_{2} < x_{1}}\sum\limits_{k=0}^{N-1}{e^{2\pi i x_{2}k/N}|k\rangle}\stackrel{I^{\otimes n+1}\otimes IQFT_{n}}{\xrightarrow{\qquad}}\\
    &|x_{1}\rangle|0\rangle_{x_{2} \geq x_{1}}|x_{2}\rangle + |x_{1}\rangle|1\rangle_{x_{2} < x_{1}}|x_{2}\rangle \xrightarrow{\text{equal}}\\
    &|x_{1}\rangle|x_{2}\rangle|0\rangle_{x_{2} < x_{1}}
\end{align}

The circuit compares the value between quantum states and preserves the states of the qubits themselves, which can be further utilized in subsequent quantum circuits based on the properties of the existing qubits.
\clearpage
\section{Modular Arithmetic Extension}\label{appendix2}
In Section IV.B, we construct a modular addition circuit depending on our proposed quantum integer comparators. It is worth noting that modulo-addition is a basic component of other modulo operations. In the following, let $N\in\mathbb{N}_{+}$ be an modulo and $n = \lceil\log_{2}N\rceil$.
we will further introduce modular multiplication and modular power components based on our modulo addition.  

\begin{figure*}[h]
\includegraphics[width=0.75\textwidth]{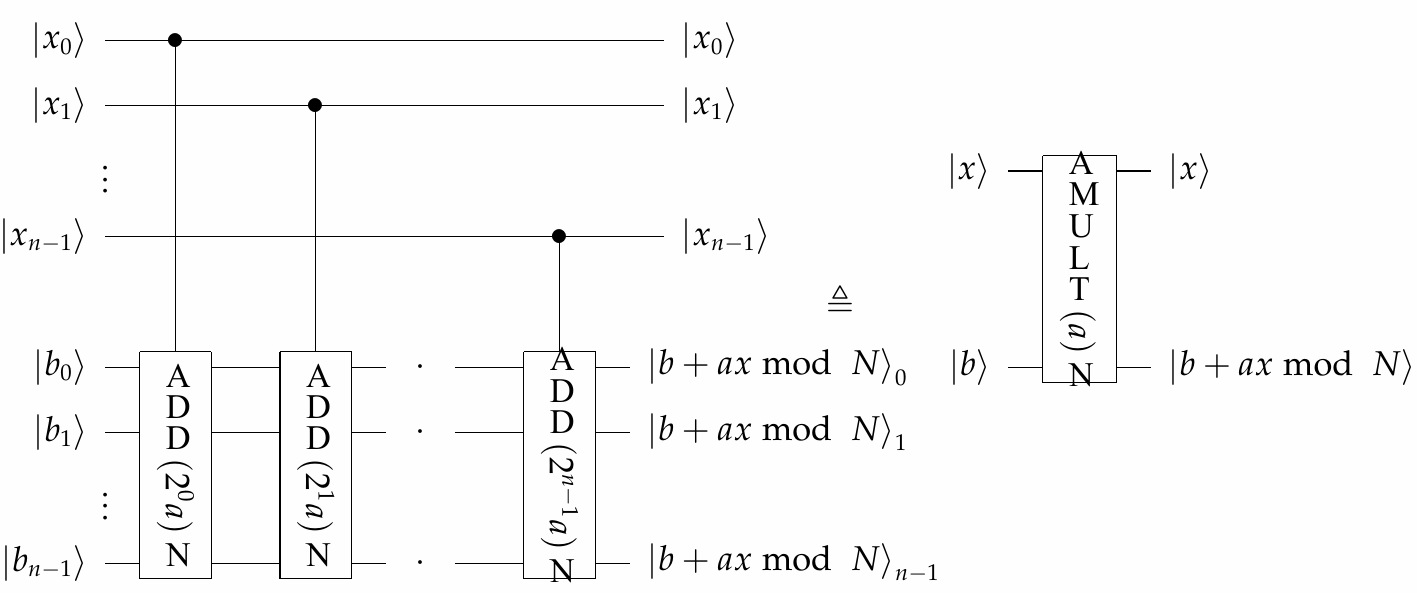}
\caption{The circuit can compute the modular addition-multiplication that $|x\rangle|b\rangle\rightarrow|x\rangle|(b+ax)\bmod N\rangle $. (Abbreviated as $\text{AMULN}$) Here, we use ADDN to represent the proposed modular addition.}
\label{fig:conmodaddmul}
\end{figure*}

If we want to construct a quantum circuit to compute modular addition-multiplication that $y+ax \bmod N$ with variables $x, y \in [0, N-1]$, we can use the following equation:
\begin{equation}
	ax\bmod N=(\ldots((2^{0}ax_0)\bmod N+2^{1}ax_{1})\bmod N+\ldots+2^{n-1}ax_{n-1})\bmod N,
\end{equation}
where $x=2^0 x_0+2^1 x_1+2^2 x_2+\cdots +2^{n-1}x_{n-1}, x_i\in \{0,1\}$.
Then, implementing modular multiplication will require $n-1$ modular additions. Thus, by combining our proposed modular addition circuit, we can implement the circuit
$|x\rangle|b\rangle\rightarrow|x\rangle|(b+ax)\bmod N\rangle $ showing in Fig.~\ref{fig:conmodaddmul}. 

\begin{figure*}[h]
\includegraphics[width=0.7\textwidth]{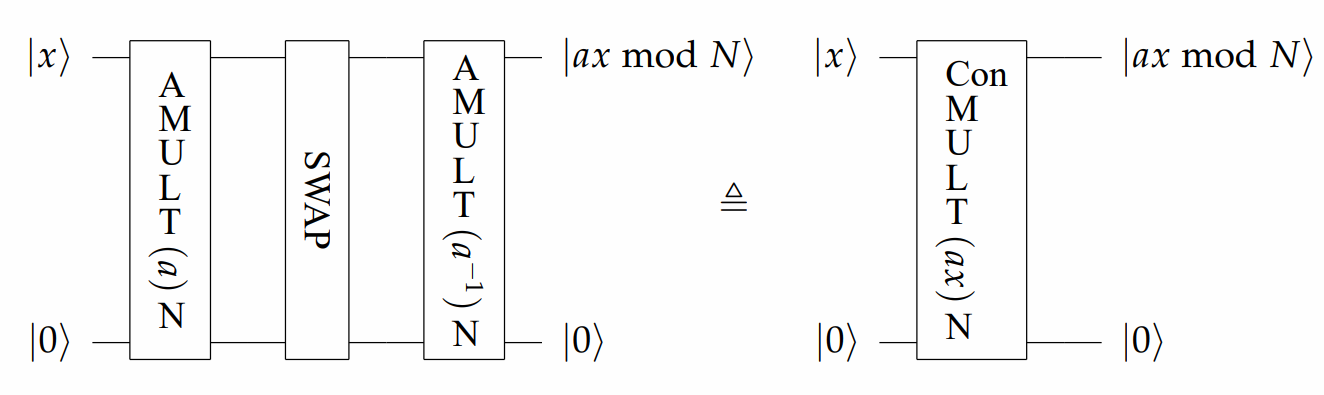}
\caption{The circuit can compute the modular multiplication that $|x\rangle|0\rangle\rightarrow|ax\bmod N\rangle|0\rangle $. (Abbreviated as $\text{ConMULTN}$)}
\label{fig:conmodmul}
\end{figure*}

Furthermore, to implement a reversible quantum circuit for modular multiplication, we use the relation 
\begin{equation}
|x\rangle|0\rangle\rightarrow|x\rangle|ax\bmod N\rangle\rightarrow |ax\bmod N\rangle|x\rangle\rightarrow |ax\bmod N\rangle|x-a^{-1}ax\bmod N\rangle=|ax\bmod N\rangle|0\rangle.
\end{equation}
Based on the relation, the modular multiplication quantum circuit can be constructed as shown in Figure \ref{fig:conmodmul}, which requires $2n+1$ qubits.

\begin{figure*}[h]
\includegraphics[width=0.6\textwidth]{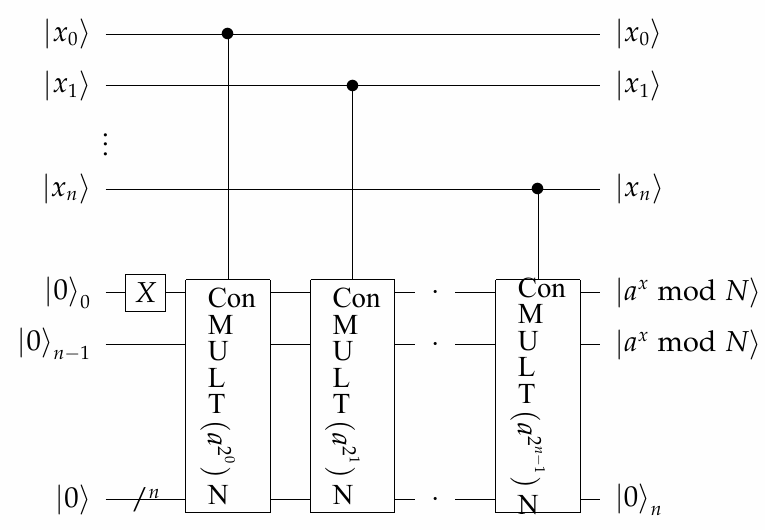}
\caption{The circuit can compute the modular exponentiation $ a^{x}\bmod N $. }
\label{fig:conmodexp}
\end{figure*}

The quantum circuit for modular exponentiation based on QFT can be constructed through the modular multiplication circuit with an additional quantum register, which is used to store the modular addition result in each iteration of controlled modular multiplication, see the following equation Eq.(\ref{eq:modexp}). The circuit is shown in Figure.\ref{fig:conmodexp}.
During the circuit implementation, three quantum registers are required, requiring $3n+1$ qubits.
\begin{equation}
	\begin{split}
		a^{x} \bmod N &= a^{2^0 x_0+2^1 x_1+2^2 x_2+\cdots +2^{n-1}x_{n-1}}\bmod N\\
		& =((a^{2^0x_0}\bmod N)\cdot (a^{2^1x_1}\bmod N)\cdots (a^{2^{n-1}x_{n-1}}\bmod N))\bmod N
	\end{split}
\end{equation}\label{eq:modexp}

\end{document}